\documentclass[twocolumn]{aastex631}

\usepackage{amsmath}
\usepackage{sistyle}

\SIthousandsep{,}

\graphicspath{{./}{figures/}}

\newcommand{\msun}{\mathrm{M_{\odot}}} 
\newcommand{\msunyr}{\msun\,{\rm yr}^{-1}} 

\newcommand{\mhe}{M_{\mathrm{He}}} 
\newcommand{\mwd}{M_{\mathrm{WD}}} 

\newcommand{\mdot}{\dot{M}} 
\newcommand{\mdotup}{\dot{M}^{+}_{\mathrm{cr}}} 
\newcommand{\mdotlow}{\dot{M}^{-}_{\mathrm{cr}}} 

\newcommand{\lsun}{\mathrm{L_{\odot}}} 
\newcommand{\Lhe}{L_{\mathrm{He}}} 
\newcommand{\Lwd}{L_{\mathrm{WD}}} 

\newcommand{\Lacc}{L_{\mathrm{acc}}}

\newcommand{\Twd}{T_{c,\mathrm{WD}}}
\newcommand{\Teffwd}{T_{\mathrm{eff,WD}}}

\newcommand{\Rhe}{R_{\mathrm{He}}} 
\newcommand{\Rwd}{R_{\mathrm{WD}}} 

\newcommand{\iniMhe}{M^{i}_{\mathrm{He}}}
\newcommand{\iniMwd}{M^{i}_{\mathrm{WD}}}
\newcommand{\iniThe}{T^{i}_{c,\mathrm{He}}}
\newcommand{\iniShe}{S^{i}_{c,\mathrm{He}}}
\newcommand{\iniTwd}{T^{i}_{c,\mathrm{WD}}}
\newcommand{\iniP}{ P^{i}_{\mathrm{orb}}}

\newcommand{\finalMwd}{M^{f}_{\mathrm{WD}}}

\newcommand{\mesa}{{\tt\string MESA}}

\newcommand{\binary}{\tt\string binary}

\newcommand{\tcond}{t_{\mathrm{cond}}}
\newcommand{\kB}{k_{\mathrm{B}}}
\newcommand{\NA}{N_{\mathrm{A}}}

\newcommand{\taum}{\tau_{\mathrm{m}}}
\newcommand{\tauth}{\tau_{\mathrm{th}}}
\newcommand{\cp}{c_{\mathrm{p}}}
\newcommand{\CV}{C_{\mathrm{V}}}
\newcommand{\taucool}{\tau_{\mathrm{cool}}}
\newcommand{\taucoolAM}{\tau_{\mathrm{cool,AM\,CVn}}}
\newcommand{\taucoolCO}{\tau_{\mathrm{cool,C/O}}}

\newcommand{\Porb}{P_{\mathrm{orb}}}

\newcommand{\mdotcrp}{\dot{M}^{+}_{\mathrm{cr}}}
\newcommand{\mdotcrm}{\dot{M}^{-}_{\mathrm{cr}}}

\newcommand{\Teffcrm}{T^{-}_{\mathrm{eff,cr}}}

\newcommand{\Teff}{T_{\mathrm{eff}}}
\newcommand{\Tstar}{T_{*}}
\newcommand{\Rtid}{R_{\mathrm{t}}}

\newcommand{\Tacc}{T_{\mathrm{acc}}}

\newcommand{\sigmasb}{\sigma_{\mathrm{sb}}}

\newcommand{\MV}{M_{\mathrm{V}}}
\newcommand{\Mg}{M_{\mathrm{g}}}

\newcommand{\Tc}{\ensuremath{T_{\rm c}}}
\newcommand{\rhoc}{\ensuremath{\rho_{\rm c}}}
\newcommand{\inipsic}{\psi^{i}_{c,\mathrm{He}}}

\newcommand{\diff}{\mathrm{d}}

\newcommand{\CO}{_{\mathrm{CO}}}
\newcommand{\he}{_{\mathrm{He}}}
\newcommand{\Mtot}{M_{\mathrm{tot}}}

\begin{document}

\title{Mass Transfer and Stellar Evolution of the White Dwarfs in AM CVn Binaries}

\author[0000-0001-9195-7390]{Tin Long Sunny Wong}
\affiliation{Department of Physics, University of California, Santa Barbara, CA 93106, USA}

\author[0000-0001-8038-6836]{Lars Bildsten}
\affiliation{Department of Physics, University of California, Santa Barbara, CA 93106, USA}
\affiliation{Kavli Institute for Theoretical Physics, University of California, Santa Barbara, CA 93106, USA}

\correspondingauthor{Tin Long Sunny Wong}
\email{tinlongsunny@ucsb.edu}

\begin{abstract}

We calculate the stellar evolution of both white dwarfs (WDs) in AM
CVn binaries with orbital periods of $\Porb \approx 5-70$
minutes. We focus on the cases where the donor starts as a $\mhe < 0.2 \, \msun$ Helium WD and the accretor is a $\mwd > 0.6 \, \msun$ WD. Using Modules for Experiments in Stellar
Astrophysics (MESA), we simultaneously evolve both WDs assuming
conservative mass transfer and angular momentum loss from
gravitational radiation. This self-consistent evolution yields the
important feedback of the properties of the donor on the mass transfer
rate, $\mdot$, as well as the thermal evolution of the accreting
WD. Consistent with earlier work, we find that the high $\mdot$'s at
early times forces an adiabatic evolution of the donor for $\Porb < 30$ minutes so that its mass-radius relation depends primarily
on its initial entropy. As the donor reaches $ \mhe \approx
0.02-0.03 \, \msun$ at $\Porb \simeq 30 $ minutes, it becomes
fully convective and could lose entropy and expand much less than
expected under further mass loss. However, we show that the lack of
reliable opacities for the donor's surface inhibit a secure prediction
for this possible cooling. 
Our calculations capture the core heating that occurs during the first $\approx 10^7$ years of accretion and continue the evolution into the phase of WD cooling that follows.  When compared to existing data for accreting WDs, as seen by Cheng and collaborators for isolated WDs, we also find that the accreting WDs are not as cool as we would expect given the amount of time they have had to cool.

\end{abstract} 


\section{Introduction}
\label{sec:intro}

AM Canum Venaticorum (AM CVn) systems are ultracompact binaries undergoing helium mass-transfer with orbital periods, $\Porb$, between 5 and 68 minutes \citep[e.g.,][]{Nather1981,2001A&A...368..939N,Nelemans2004,Ramsay2018}. Their orbital evolution is dominated by loss of angular momentum from gravitational wave radiation \citep[e.g.,][]{1979AcA....29..665T,1981NInfo..49....3T}, and the gravitational wave signal from AM CVn systems should be detectable by missions such as the Laser Interferometer Space Antenna \citep[LISA;][]{Amaro2013,Amaro2017,Nelemans2004,Kremer2017,Breivik2018}. The study of AM CVn systems is of interest for several reasons. Their orbital evolution can help constrain the efficiency of tides in degenerate stars \citep[e.g.,][]{Piro2019}. Unstable helium burning is expected to occur in some AM CVn systems, leading to  helium novae \citep[e.g.,][]{Ashok2003}, and in the case of dynamical helium-burning, faint thermonuclear ``.Ia'' supernovae \citep[][]{Bildsten2007,Shen2009}. Accretion is expected to occur via an accretion disk \citep[except at shortest periods where direct-impact accretion occurs; e.g., ][]{2004MNRAS.350..113M}, which allows AM CVn systems to be good testing grounds for the theory of nearly pure helium accretion disks \citep[e.g.,][]{Kotko2012}. 

There are three possible channels for the formation of AM CVn systems. In the helium white dwarf (He WD) donor scenario, the initial donor is a degenerate, low-mass ($\approx 0.1 - 0.2 \, \msun$) helium white dwarf \citep[e.g.,][]{Deloye2007}. In the helium star (He star) donor scenario, the donor starts mass-transfer as a non-degenerate, helium-burning star with masses ranging from $\approx 0.3$ to $0.7 \, \msun$ \citep[e.g.,][]{1991ApJ...370..615I,Yungelson2008}. Note that at long periods $\Porb \gtrsim 40 \, \mathrm{min}$, the thermal properties of the He WD and He star donors are predicted to converge \citep[][]{Deloye2007,Yungelson2008}. The two remain distinguishable by their compositions, with the He star donors expected to contain products of helium-burning. In the cataclysmic variable (CV) donor channel \citep[e.g.,][]{Podsiadlowski2003}, the donor starts as a $\approx 1 \, \msun$ star, initiates hydrogen-rich mass transfer around the end of core hydrogen-burning, and eventually loses its hydrogen-rich envelope. In all three channels, the accretor is often assumed to be a $0.6 - 1.0 \, \msun$ carbon-oxygen white dwarf (CO WD).

In this work, we revisit the calculations by \cite{B06} on the thermal evolution of the CO WD accretor in the context of the He WD donor channel. In Section \ref{sec:computational setup}, we describe our computational setup and construction of initial stellar models for both the donor and the accretor. We evolve donor models of various initial central entropies under mass transfer to a point-mass, and describe their thermal evolution in Section \ref{sec:donor}. We show that the mass-radius relation of the donor, as well as the time evolution of $\Porb$, depends on their initial central entropy and whether they can cool. In Section \ref{sec:heating}, we evolve the CO WD accretor along with the donor and the binary orbit. Our results agree with \cite{B06} but are more consistent as our models track the time-dependent cooling of the donor. 
At high mass transfer rates, $\mdot$, near period minimum, the accretor is reheated due to accretion; as the orbit widens and $\mdot$ drops, the accretor luminosity eventually becomes just that of a cooling WD. We compute synthetic color-magnitude diagrams for our accretor models for comparisons with observations in Section \ref{sec:comparison to observations}. We show that observed AM CVn systems appear bluer and brighter than expected, which corresponds to a younger WD cooling age. We conclude in Section \ref{sec:conclusion}.

In the following, we denote the accretor and the donor by the subscripts WD and He. Quantities at the center of the stars are denoted by the subscript $c$. Initial and final quantities are denoted by the superscripts $i$ and $f$. Age since initiation of mass transfer will be used synonymously with age. 


\section{Computational Setup and Initial Stellar Model Construction}
\label{sec:computational setup}

To obtain realistic, time-dependent mass transfer histories for AM CVn binaries, we self-consistently evolve a He WD donor and a CO WD accretor using the $\binary$ capability of the stellar evolution instrument Modules for Experiments in Stellar Astrophysics ($\mesa$ version 12778; \citealt{2011ApJS..192....3P,2013ApJS..208....4P,2015ApJS..220...15P,2018ApJS..234...34P,2019ApJS..243...10P}). 
The equation of state (EOS) adopted for the WDs in this study is a blend among OPAL \citep{Rogers2002} and SCVH \citep{Saumon1995} at low densities ($\log_{10} (\rho / \mathrm{g} \, \mathrm{cm}^{-3}) \lesssim 2.9$), and PC at high densities \citep[][$\log_{10} (\rho / \mathrm{g} \, \mathrm{cm}^{-3}) \gtrsim 2.9$]{Potekhin2010}. Radiative opacities, where available, are taken from OPAL \citep{Iglesias1993,
Iglesias1996} at $\log_{10} (T/\mathrm{K}) \gtrsim 3.8$ and \cite{2005ApJ...623..585F} at $\log_{10} (T/\mathrm{K}) \lesssim 3.8$, and extrapolated otherwise (see Section \ref{sec:Donor Thermal Evolution Uncertainties Under Mass Transfer}). Electron conductive opacities are from \cite{2007ApJ...661.1094C} with corrections from \cite{2020ApJ...899...46B} for He under moderate coupling and moderate degeneracy. The latter corrections are implemented by modifying our copy of the $\mesa$ source code, but will be included in future releases of $\mesa$. Our nuclear net includes the NCO reaction chain \citep{2017ApJ...845...97B} which may impact the helium shell thickness for unstable burning on the accretor. The models presented in this work differ slightly from an earlier version of models presented in \cite{vanRoestel2021}. The latter adopted the HELM EOS \citep[][]{Timmes2000} during the early phases of mass transfer for the He WD donor, and included no electron conduction opacity correction from \cite{2020ApJ...899...46B} for both WDs. However, difference in quantities like luminosity is insignificant (of order 10\%). Our $\mesa$ input files are available at Zenodo\footnote{\url{https://doi.org/10.5281/zenodo.5532940}}. 

The He WD models are made using a modified version of the $\tt \string make\_he\_wd$ test suite, where a 1.5 $\msun$ star with metallicity $Z=0.02$ is evolved from the zero-age main sequence (ZAMS) to the formation of an inert helium core of mass $0.15-0.18 \, \msun$. We then strip off the envelope, and allow the bare He core to cool to the desired central entropy. Extremely low mass (ELM) He WDs are expected to have a hydrogen envelope of mass $\approx 10^{-3} \, \msun$ \citep[e.g.,][]{2016A&A...595A..35I}. When accreted onto the CO WD the hydrogen may undergo unstable nuclear burning \citep[e.g.,][]{2012ApJ...758...64K}, which is computationally expensive to follow and has no thermal impact on the accretor, hence we neglect it in our study. We leave the He WD with almost no hydrogen envelope ($M_{\mathrm{H}} \lesssim 10^{-20} \, \msun$), which only impacts the initial mass transfer phase, but has no impact on the mass transfer at late times \citep[e.g.,][]{2012ApJ...758...64K}.

The He WD models have initial specific central entropies of $\iniShe/\NA \kB = 2.23, 2.61, 3.07, 3.52$, where $\NA$ is Avogadro's number.
In \cite{Deloye2007}, the donor models are labeled by their initial central degeneracy parameter, $\inipsic = E_{\mathrm{F,c}} / \kB \Tc \approx \rhoc / ( 1.2 \times 10^{-8} \Tc^{3/2} ) \, \mathrm{K}^{3/2} \mathrm{cm}^{3} \mathrm{g}^{-1} $, where $\Tc$, $\rhoc$ and $E_{\mathrm{F,c}}$ are the initial temperature, density and electron Fermi energy at the center of the He WD. Our models correspond to $\log_{10} (\inipsic) = 2.89, 2.39, 1.82, 1.34$ respectively. 
Our models are chosen to be similar in range of $\inipsic$ to those in \cite{Deloye2007}. \citealt{Deloye2007} determined the distribution of $\inipsic$ at contact (see their Fig. 3) based on post-common envelope conditions of the surviving double WD systems in the population synthesis study of \cite{2001A&A...368..939N}, and modeled the mass transfer of He WDs with $\log_{10}(\inipsic) \approx 1.1 - 3.5 $. Our coldest model is less degenerate than theirs due to inadequate EOS coverage in $\mesa$ (see below).

The CO WD models are similarly made using a modified version of the $\tt \string make\_co\_wd$ test suite. We evolve a ZAMS model of mass $3-6 \, \msun$ with metallicity $Z=0.02$ up to helium shell burning and the formation of a CO core of the desired mass. We then remove the envelope and allow the CO core to cool to the desired starting core temperature.

The He WD and CO WD models are then taken as initial models in $\mesa$ $\binary$ which self-consistently evolves the stellar structures of both stars and the orbital parameters. We assume fully conservative mass transfer at a rate $\mdot$ and model it using the Ritter scheme \citep{1988A&A...202...93R}. We assume that orbital angular momentum loss is only due to the emission of gravitational waves. While the accretor may be significantly spun up by the accreted material at the expense of orbital angular momentum \citep{2004MNRAS.350..113M,2007ApJ...655.1010G}, we choose to model both stars as non-rotating as a first approximation. Tides may play an important role in synchronizing both components, which helps stabilize the orbit, and in impacting their thermal evolution through tidal dissipation \citep[e.g.,][]{2004MNRAS.350..113M,2012MNRAS.421..426F,2014MNRAS.444.3488F}. 
Observations show that at long periods ($\Porb = 46 \, \mathrm{min}$ for GP Com and $65 \, \mathrm{min}$ for V396 Hya) the accretor is rotating far slower than critical \citep{2016MNRAS.457.1828K}. 

The binary simulations begin with an initial orbital period of $\approx 20$ minutes with both components detached. The orbit decays due to gravitational wave radiation and the He WD eventually fills its Roche lobe. We follow the evolution of the system until the donor has cooled down sufficiently that $\log Q = \log \rho - 2 \log T + 12 \geqslant 5 $ in any of the cells. This choice was made because for $\log Q \geqslant 5$, $\mesa$ switches to an ideal gas equation of state which affects the donor mass-radius relation and subsequently the orbital evolution. This typically occurs only when $\Porb \gtrsim 60 \, \mathrm{min} $, allowing us to compare our predictions with all known AM CVn systems. Future work with $\mesa$ incorporating the EOS's of \cite{2019ApJ...872...51C} and \cite{2021ApJ...913...72J} can probe the further evolution of the donor at even lower temperatures with better numerical accuracy. 

For the accretor, we set $\tt tau\_factor = 30$ for numerical convenience, which places the outermost cell at an optical depth of $\tau = 2/3 \times 30 = 20$. We also opted to adopt a grey Eddington $T-\tau$ relation instead of interpolating from the DB tables provided by Odette Toloza and Detlev Koester. The level of discrepancy in the accretor effective temperature for these two approaches is $\lesssim 10\%$. We allow the latent heat of crystallization to be released at a Coulomb coupling parameter $\Gamma = \langle Z^{5/3}_{i} \rangle e^{2} / a_{e} \kB T$ between 225 and 235 \citep[see equation 1 of][]{Bauer2020}, where $\langle Z^{5/3}_{i} \rangle$ is an average of $Z^{5/3}_{i}$ over all ion species with charge $Z_{i}$, and $a_{e} = ( 3 / 4 \pi n_{e} )^{1/3} $ is the electron separation. 

As we show in Section \ref{sec:donor}, the initial thermal properties and subsequent thermal evolution of the donor are important in setting the binary orbital evolution and $\mdot$ history. 
Therefore, for each $\iniMhe = 0.15 \, \msun$ model of fixed $\iniShe$, we generate an adiabatic mass-radius relation, by stripping the $ 0.15 \, \msun$ He WD model to the desired mass via the $\tt \string relax\_mass$ control in $\tt \string MESA$. The rate of mass change for this construction is chosen such that the core of the model evolves adiabatically. 
We obtain the cold WD (fully degenerate) limit similarly by stripping the $ 0.15 \, \msun$ He WD models to the desired mass, and subsequently allowing them to cool to a central temperature of $T_{\mathrm{c,He}} = 10^{5} \, \mathrm{K}$. To avoid the $\log Q \geqslant 5 $ stopping condition, we irradiate the surface layer of the He WD model. This is done in part for numerical convenience, but is also expected for realistic AM CVn systems where, on geometrical grounds, the accreting WD should strongly irradiate the donor.

For comparison, we also evolved a $\iniMhe = 0.35 \, \msun$ He star model taken from \cite{Brooks2015}. We chose $\iniMwd = 0.5 \, \msun$ and $\iniP = 20 \, \mathrm{min}$, which corresponds to the first model shown in Table 1 of their paper and of \cite{Yungelson2008}. 
We also obtained an adiabatic mass-radius relation for the He star using the $\tt \string relax\_mass$ control described above. The initial model for this construction is taken from the binary simulation when core He burning is quenched in the He star and its mass decreases to $0.2 \, \msun$.


\section{Donor Evolution}
\label{sec:donor}

\begin{figure}
\fig{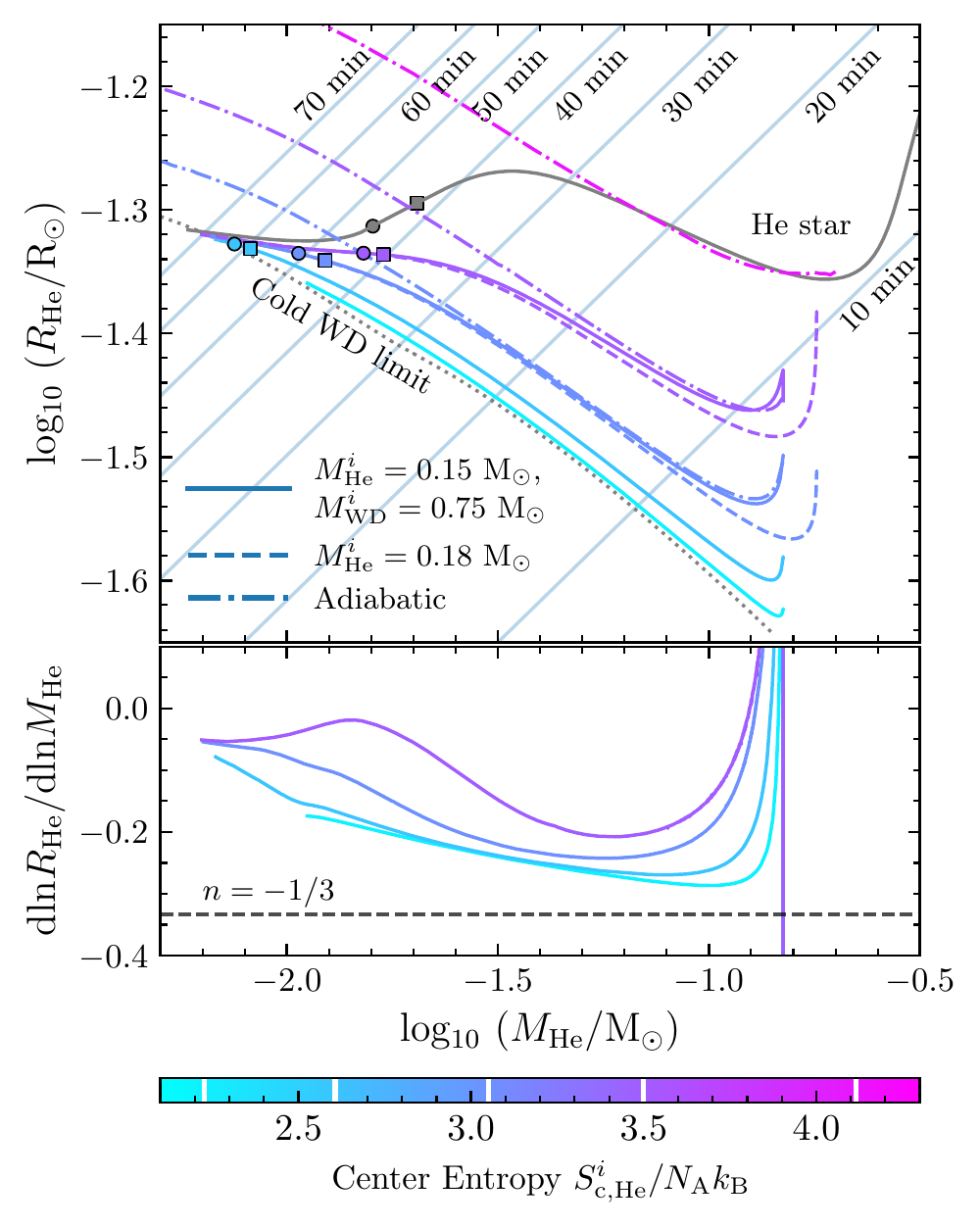}{ 0.49 \textwidth}{}
\caption{ Donor mass-radius relation (top panel) and the associated power-law index $n = \diff \ln \Rhe / \diff \ln \mhe$ (bottom panel), color-coded by their initial central specific entropy, $\iniShe$ (labeled by white lines in the colorbar). We take $\iniMhe = 0.15 \, \msun$ (solid lines) or $0.18 \, \msun $ (dashed lines), both with $\iniMwd= 0.75 \, \msun$. Square symbols indicate where the donor thermal timescale, $\tauth = \int \cp T \diff m / \Lhe$, first becomes smaller than the mass transfer timescale, $\taum = \mhe/\mdot$. Circle symbols indicate where the donor first becomes fully convective. For the $\iniMhe = 0.15 \, \msun$ models, we show the adiabatic (dot-dashed lines) and fully degenerate (taken as $T_{\mathrm{c,He}} = 10^{5} \, \mathrm{K} $; dotted grey line) mass-radius relations. We also show a He star model (grey solid line) taken from \cite{Brooks2015}, and the corresponding adiabatic mass-radius relation starting from $ 0.2 \, \msun$. 
For comparison, we show lines of constant $\Porb$ (light blue lines) using the mean density - period relation for a Roche-lobe filling object. 
\label{fig:donor_mass_radius}
}
\end{figure}

\begin{figure}
\fig{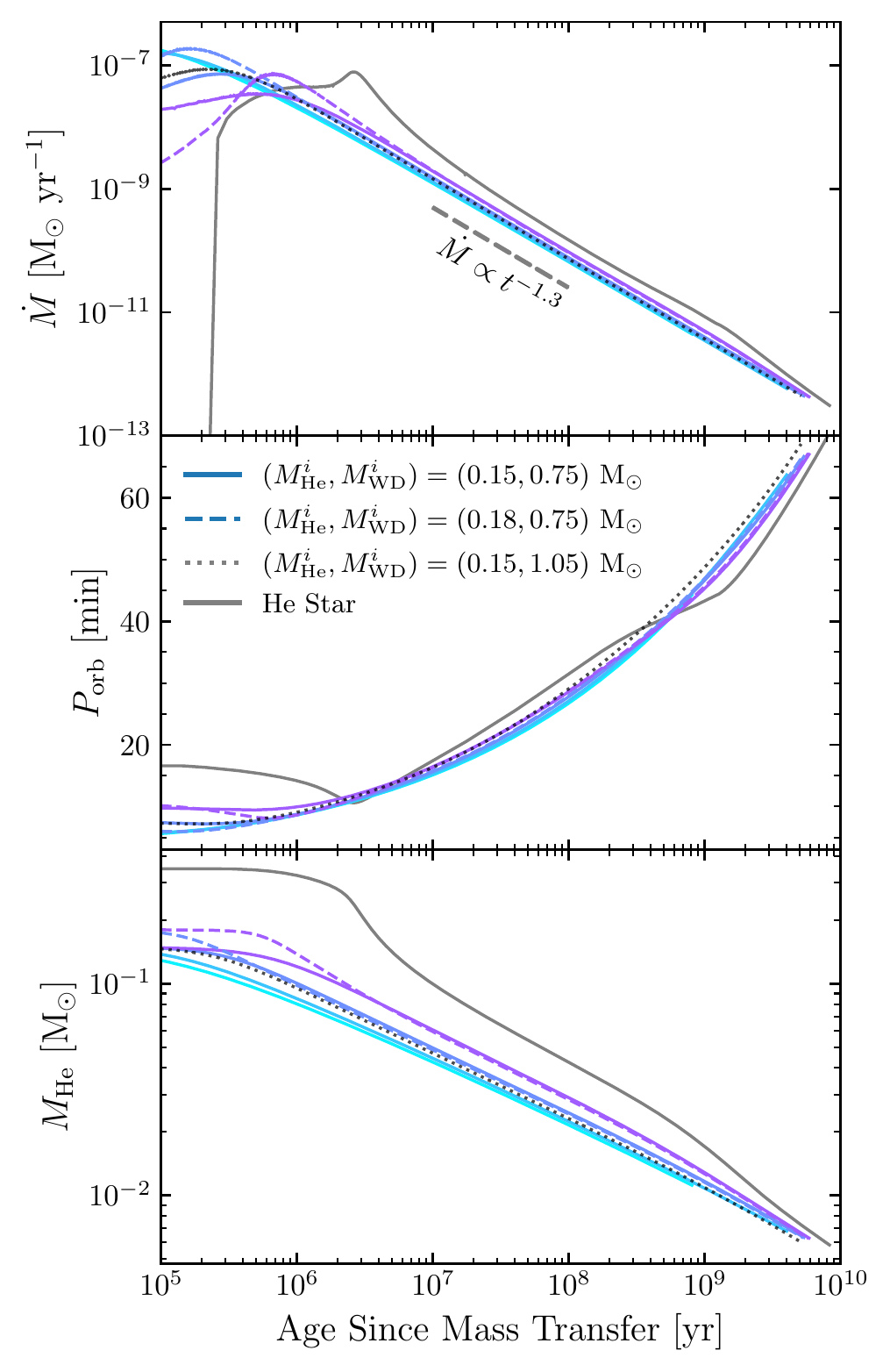}{ 0.49 \textwidth}{}
\caption{ Time evolution of the mass transfer rate, $\mdot$ (top panel), orbital period, $\Porb$ (middle panel), and donor mass, $\mhe$ (bottom panel), for various initial donor central specific entropies, $\iniShe$. The colored solid (grey dot-dashed) lines are for $\iniMwd=0.75 \, (1.05) \, \msun$ and $\iniMhe = 0.15 \, \msun$, the colored dashed lines are for $\iniMwd=0.75 \, \msun$ and $\iniMhe = 0.18 \, \msun$, and the solid grey line is for the He star model. 
For comparison, we show $\mdot \propto t^{-1.3}$ \citep[dashed grey line; ][]{B06} as analytically expected. 
\label{fig:donor_time_comparison}
}
\end{figure}

The mass-transfer rate, $\mdot$, determines whether the accretor is heating or cooling, and sets the evolution of binary parameters \citep[e.g., orbital period, $\Porb$; ][]{B06}. It depends on the donor's mass-radius relation, which in turn is set by its thermal evolution \citep{Deloye2007}. Thus we start by exploring the effect of varying the donor's initial central specific entropy, $\iniShe$, on $\mdot$ and the age-period relation. We run $\mesa$ models for initial donor masses of $\iniMhe = 0.15 \, \msun$ or $ 0.18 \, \msun$ with $\iniShe/\NA \kB = 2.23, 2.61, 3.07, 3.52$.  
As we are first exploring the $\mdot(t)$ and $\Porb(t)$ evolution, we model the accretor as a point-mass with an initial $\iniMwd = 0.75 \, \msun$. 
We show the effect of an initially more massive accretor by another model with $\iniShe/\NA \kB = 3.07$ and $\iniMwd=1.05 \, \msun$. 

\subsection{Donor Mass-Radius Relation}
\label{sec:donor mass-radius relation}

The resulting donor's mass-radius relations are shown in the top panel of Figure \ref{fig:donor_mass_radius}, and the associated power-law index $n = \diff \ln \Rhe/ \diff \ln \mhe$ in the bottom panel. Initially, the donor contracts ($n > 0$) as the outermost radiative layer is stripped off (see \citealt{Deloye2007} and \citealt{2012ApJ...758...64K} for in-depth discussions). Eventually, the underlying layers drive an expansion in radius ($n < 0$), with the lowest entropy model closest to $n = - 1/3$ as expected for a fully degenerate object. In general, the less degenerate, higher entropy models show a more positive $n$. As $\mhe$ decreases, the power-law slope becomes more positive, in part because Coulomb effects become more important \citep{Deloye2003}. 

For less degenerate donors (especially the $\iniShe/\NA \kB = 3.52$ donor), $n$ becomes more positive when they start to cool. Initially, the donors evolve adiabatically as the mass change timescale, $\taum = \mhe/\mdot$, is much shorter than the thermal timescale, $\tauth = \int \cp T \diff m / \Lhe$. This is confirmed by the agreement of the $R(M)$ relations from the binary calculations (solid lines) and the adiabatic models (dot-dashed lines) in Figure \ref{fig:donor_mass_radius}, below $\Porb \approx 30 \, \mathrm{min}$ when $\taum \ll \tauth$. 
As $\mdot$ drops with time, eventually $\tauth \lesssim \taum$ ($\tauth = \taum$ is labeled by square symbols). Then the donor becomes fully convective (labeled by circle symbols) and starts to cool \citep{Deloye2007}, as we discuss in the end of this section. Therefore, less degenerate donors start to converge to the mass-radius relation of a fully degenerate He WD, as seen in Figure \ref{fig:donor_mass_radius} \citep[see also Fig. 8 of][]{Deloye2007}. 

The mass-radius relation depends only slightly on $\iniMhe$, as seen in Figure \ref{fig:donor_mass_radius}. At a given $\iniShe$, the difference between the $\iniMhe = 0.15 \, \msun$ (solid lines) and $0.18 \, \msun$ (dashed lines) tracks is only evident at larger $\mhe$ ($\Porb \lesssim 20 \, \mathrm{min}$), and is negligible at small $\mhe$ especially when donor cooling happens. 

\subsection{Mass Accretion Rate Histories}
\label{sec: mass accretion rate histories}

We show the time evolution of $\mdot$ in the top panel of Figure \ref{fig:donor_time_comparison}. As the He WD donor fills its Roche lobe, the mass transfer rate rapidly rises to $\sim 10^{-7} - 10^{-8}$ $\msunyr$. 
All $\mdot$'s peak within $10^{6}$ yrs after mass transfer initiates, with more degenerate donors having higher $\mdot$ due to their smaller radii \citep[e.g., ][]{2012ApJ...758...64K}. Afterwards, $\mdot \propto t^{-1.3}$ as analytically expected \citep{B06}. At a fixed age, $\mdot$ is lower for more degenerate donors. The ordering of $\mdot$ with donor degeneracy both at peak $\mdot$ and in the power-law phase agrees with \citealt{Deloye2007} (their Fig. 6). The top panel of Figure \ref{fig:donor_time_comparison} also shows that $\mdot$ does not vary significantly with $\iniMwd$, as shown by the models with $\iniMwd=0.75 \, \msun$ and $1.05 \, \msun$ at $\iniShe/\NA \kB = 3.07$. 

The time evolution of $\Porb$ is tied intimately to the donor's mass-radius relation and its thermal evolution, as shown in the middle panel of Figure \ref{fig:donor_time_comparison}. Less degenerate donors have a larger $\Porb$ at period minimum, but $\Porb$ evolution with time slows down after an age of $\approx 10^{8}$ yrs, so that at the end, more degenerate donors catch up and have larger $\Porb$ at a fixed age. For the less degenerate donors, the reason for their slower time evolution is that they eventually cool when $\tauth \lesssim \taum$. 

Due to increased orbital angular momentum loss from gravitational wave emission, a higher total mass also speeds up the $\Porb$ evolution with time. 
This is more clearly shown by the pair of $\iniShe/\NA \kB = 3.07$ models with different $\iniMwd$, and barely observable in the pairs of models with different $\iniMhe$. However, a higher total mass increases the likelihood of a helium flash being triggered on the accretor, since less accumulated mass on the accretor is required given a higher $\mdot$ (a higher $\iniMhe$ leads to a higher peak $\mdot$; see top panel of Fig \ref{fig:donor_time_comparison}) and a higher $\iniMwd$ \citep[e.g., ][]{2017ApJ...845...97B}. Indeed, our calculations that simultaneously evolve the stellar structure of the accretor show that the $\mwd = 0.75 \, \msun$ accretor undergoes a helium flash for the $\iniMhe = 0.18 \, \msun $, $\iniShe/\NA \kB = 3.07$ model at $\Porb \approx 6 \, \mathrm{min}$, while the $\iniShe/\NA \kB = 3.52$ model avoids a helium flash due to the countering effect of a higher central entropy. 

\begin{figure}
\fig{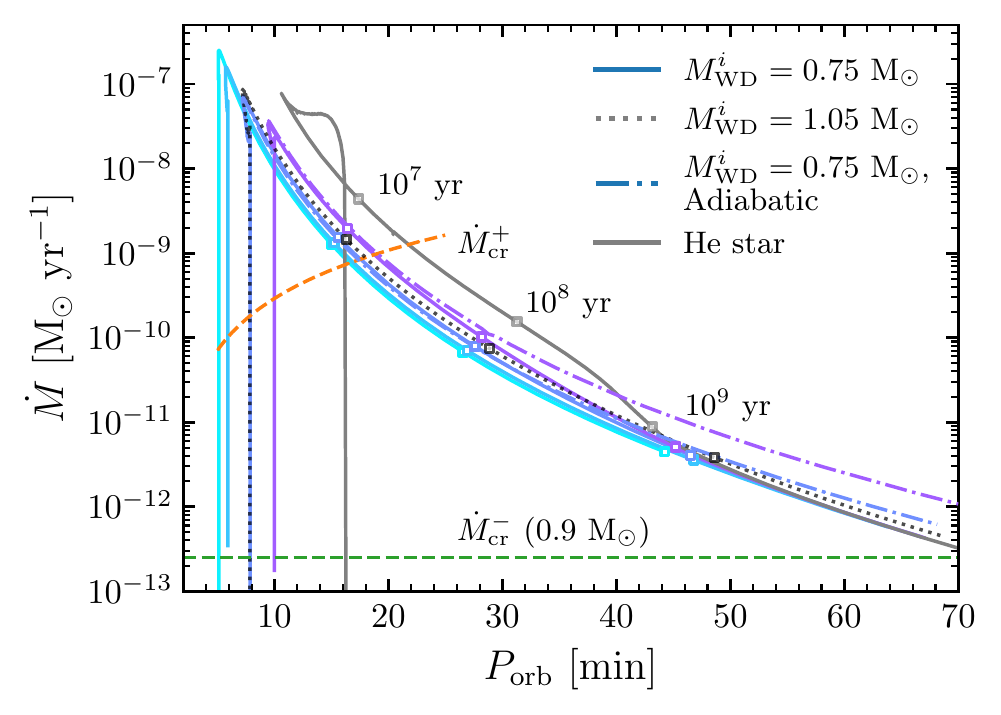}{ 0.49 \textwidth}{}
\caption{ Mass transfer rates, $\mdot$, as a function of orbital period, $\Porb$. The models shown are the same as in Figure \ref{fig:donor_time_comparison}, with the addition of semi-analytical lines for adiabatic donor evolution (dot-dashed lines). The open circles denote fixed ages since initiation of mass transfer. For comparison we show the theoretical critical mass transfer rates from the disk instability model \citep{Kotko2012}, $\mdotup$ ($\mdotlow$) with an orange (green) dashed line. The helium equivalent of dwarf novae are expected to occur between these two limits (see Appendix \ref{sec:accretion luminosity}). 
\label{fig:pm_mdot_Porb}}
\end{figure}

We show the evolution of $\mdot$ with $\Porb$ in Figure \ref{fig:pm_mdot_Porb}. For $\Porb < 40$ min, an initially hotter donor gives a higher $\mdot$ at fixed $\Porb$. But starting at $\Porb \approx 30 $ min, the $\mdot$ for the initially hotter donors start to converge to that for cold donors, reflecting the eventual cooling of the donors to a fully degenerate configuration. During this cooling, $\mdot$ drops more sharply with $\Porb$. As analytically derived by \cite{2015ApJ...803...19C}, $\mdot \propto \Porb^{\xi}$, where $\xi = 4(5-6n)/3(3n-1)$. Initially as $n \approx - 1/3$, $\xi \approx -4.67$, but during cooling, $n \approx 0$ and $\xi \approx -6.67$. Thus, the sharp drop in $\mdot$ with $\Porb$ is consistent with analytical expectations. Even though hotter donors are at longer $\Porb$ at an age of $10^{8}$ yrs (due to a larger minimum $\Porb$), this ordering reverses before an age of $10^{9}$ yrs. 

For comparison, we semi-analytically derive $\mdot - \Porb$ relations for an adiabatic donor evolution (dot-dashed lines in Figure \ref{fig:pm_mdot_Porb}). Given an adiabatic mass-radius relation from $\mesa$, we numerically integrate the binary orbital parameters ($\mhe,\mwd$ and binary separation $a$) assuming conservative mass transfer and orbital angular momentum loss solely due to gravitational waves \citep[see eqns 1 \& 2 in][]{Brooks2015}. The initial values are taken to be when $\mhe$ decreases to $ 0.14 \, \msun $ in the $\mesa$ binary run with $\iniMhe = 0.15 \, \msun$ and $\iniMwd = 0.75 \, \msun$, which correspond roughly to the moment of peak $\mdot$ in Figure \ref{fig:pm_mdot_Porb}. The excellent overlap between the solid and dot-dashed lines in Figure \ref{fig:pm_mdot_Porb} at $\Porb \lesssim 30 \, \mathrm{min}$ illustrates the initially adiabatic donor evolution. 
The resulting $\iniShe/\NA\kB = 3.52$ adiabatic line corresponds approximately to the ``hot'' donor lines in Fig 2 of \cite{B06} who assumed adiabatic donor evolution as well. Since our models eventually cool, they track closely ``cool'' models regardless of $\iniShe$ once $\Porb \gtrsim 40 - 50 \, \mathrm{min}$.

\subsection{Donor Thermal Evolution Uncertainties Under Mass Transfer}
\label{sec:Donor Thermal Evolution Uncertainties Under Mass Transfer}

The impact of the thermal evolution of the donor is substantial for the observed $\mdot(\Porb)$ and $\Porb(t)$ relations and so deserves some additional scrutiny. This evolution is set by both the initial entropy and whether entropy can be lost as mass is transferred, which we now discuss. 

The donor can only cool when its thermal timescale, $\tauth$, is comparable to or shorter than the mass transfer timescale $\taum$. 
The former is inversely proportional to the donor luminosity, $\Lhe$, so lowering $\Lhe$ may delay the cooling of the donor, and speed up the evolution of $\Porb$ with age. 

\begin{figure*} 
\fig{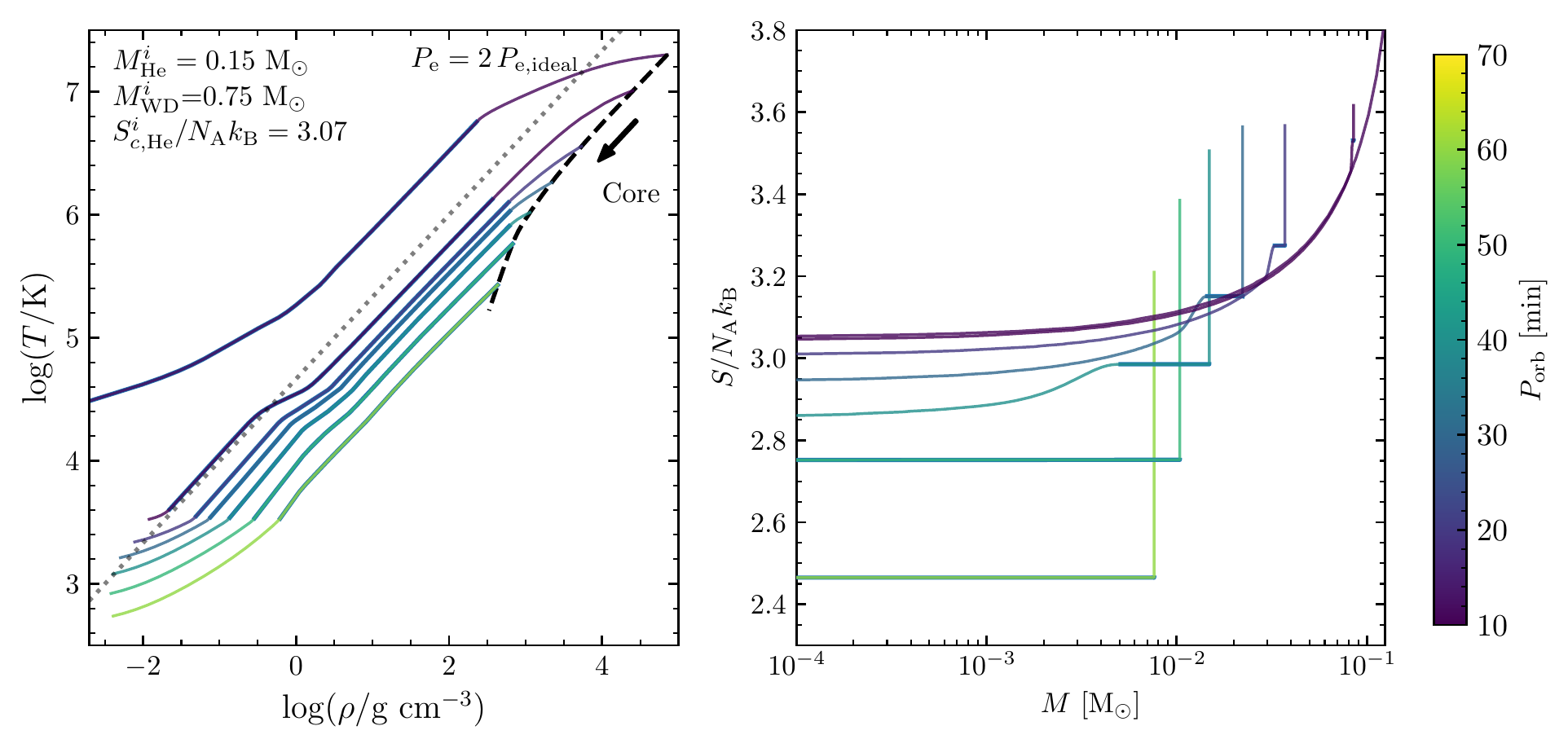}{ \textwidth }{}
\caption{ Density-temperature profiles of the donor (left panel), and entropy profiles of the donor (right panel), both color-coded by $\Porb$. Convection zones are labeled by thick blue lines. All profiles range from $\Porb = 10$ to 60 min in 10 min intervals, and we also include the initial model ($\Porb \approx 20$ min). 
\label{fig:Donor_TRho_Evolution}}
\end{figure*}

The thermal evolution of the donor with $\iniMwd=0.75 \, \msun$ and $\iniShe/ \NA \kB = 3.07$ is shown in Figure \ref{fig:Donor_TRho_Evolution}. The left panel shows a sequence of temperature-density profiles color-coded by $\Porb$, and the right panel shows the entropy profiles of the same models. The black dashed line in the left panel shows the evolution of the core in $T-\rho$ space. At $\Porb \lesssim 30 \, \mathrm{min}$, the core evolves adiabatically as $\taum \lesssim \tauth$, and starts to cool for $\Porb \gtrsim 30 \, \mathrm{min} $. This is corroborated by the entropy profiles, which show that the central entropy is roughly constant for $\Porb \lesssim 30 \, \mathrm{min}$ and only drops due to cooling at longer periods. 
This decrease of central entropy happens even before the donor becomes fully convective. Due to radiative diffusion, heat is always lost from the center, i.e., $\epsilon_{\mathrm{grav},c} \equiv - T_{c} ( \diff S_{c} / \diff t ) > 0$. The change in central entropy becomes apparent when $\Porb \gtrsim 30 \, \mathrm{min}$ as the timescale increases from $\sim 10^{7} \, \mathrm{yr}$ to $\sim 10^{8}- 10^{9} \, \mathrm{yr}$ and $\tauth$ eventually becomes comparable to, or shorter than, $\taum$. However, becoming fully convective accelerates the loss of entropy from the center, the occurrence of which we now describe. 

Before the donor becomes fully convective and cools, it has an outer convective zone (blue thick lines in Figure \ref{fig:Donor_TRho_Evolution}), and a radiative core.  The outgoing luminosity in the thin radiative surface is set by $L/\mhe = (64 \pi G / 3) ( \sigmasb T^{4} / \kappa P ) \nabla_{\mathrm{ad}} \propto 1/\kappa$, at the top of the convective boundary \citep{2006ApJ...650..394A}. 
Thus, the cooling luminosity of the donor is inversely proportional to the opacity at the top radiative-convective boundary for a fully convective star. At $\Porb \gtrsim 20 \, \mathrm{min}$, this is at $\log_{10} (\rho / \mathrm{g} \, \mathrm{cm}^{-3} ) \approx 0$ and $\log_{10} (T/ \mathrm{K}) \lesssim 4$. However, there are no radiative opacity tables currently at this range, and instead $\mesa$ extrapolates from the \cite{2005ApJ...623..585F} low-temperature opacity tables at the same $\log_{10} T$. We defer investigation of the opacity to future studies, and note that a higher opacity at the top radiative-convective boundary can delay cooling of the donor and affect the age-period relation. We also note that irradiation can play an important role in keeping the top radiative-convective boundary at a higher $T$ (and potentially $\kappa$), but this is subject to the same opacity uncertainties described above. 

The same uncertainty in the donor cooling physics exists for the He star donors at late times. From table 1 of \cite{Yungelson2008}, the typical age and $\Porb$ for $\iniMhe = 0.35 - 0.4 \, \msun$ are a few hundred Myrs and $\approx 40$ min at the end of their calculations. If we take the corresponding binary parameters, and integrate forward taking $n = -0.16$, we get an age since mass transfer between $\approx 1.5$ and $3$ Gyrs at $\Porb = 65$ min. These are lower limits nonetheless, since $n = -0.16$ is fitted for $\Porb \approx 10 - 35$ min. Eventually the He star donors should start to cool \citep{Deloye2007}, as hinted by $n \gtrsim 0$ at low masses for the mass-radius relation in Fig. 4 of \cite{Yungelson2008}. 
Indeed, this is explicitly shown by our He star model in the middle panel of Figure \ref{fig:donor_time_comparison}, where the age increases significantly around $\Porb \approx 40 \, \mathrm{min}$ as the He star starts to cool (see also Figure \ref{fig:donor_mass_radius}).

\bigskip

To summarize, we have quantified the scatter in the age-$\Porb$ relation considering all He WD models with donor cooling. In the middle panel of Figure \ref{fig:donor_time_comparison}, the range of ages at which $\Porb \approx 65$ min is $4 - 5 $ Gyrs. 
The importance of the period-age relation, itself tied closely to the thermal properties of the donor, will become clear once we discuss the heating and cooling of the accreting WD in Section \ref{sec:heating}.


\section{Heating and Cooling of the Accreting WD}
\label{sec:heating}

 \begin{figure*}[t!]
\gridline{
\fig{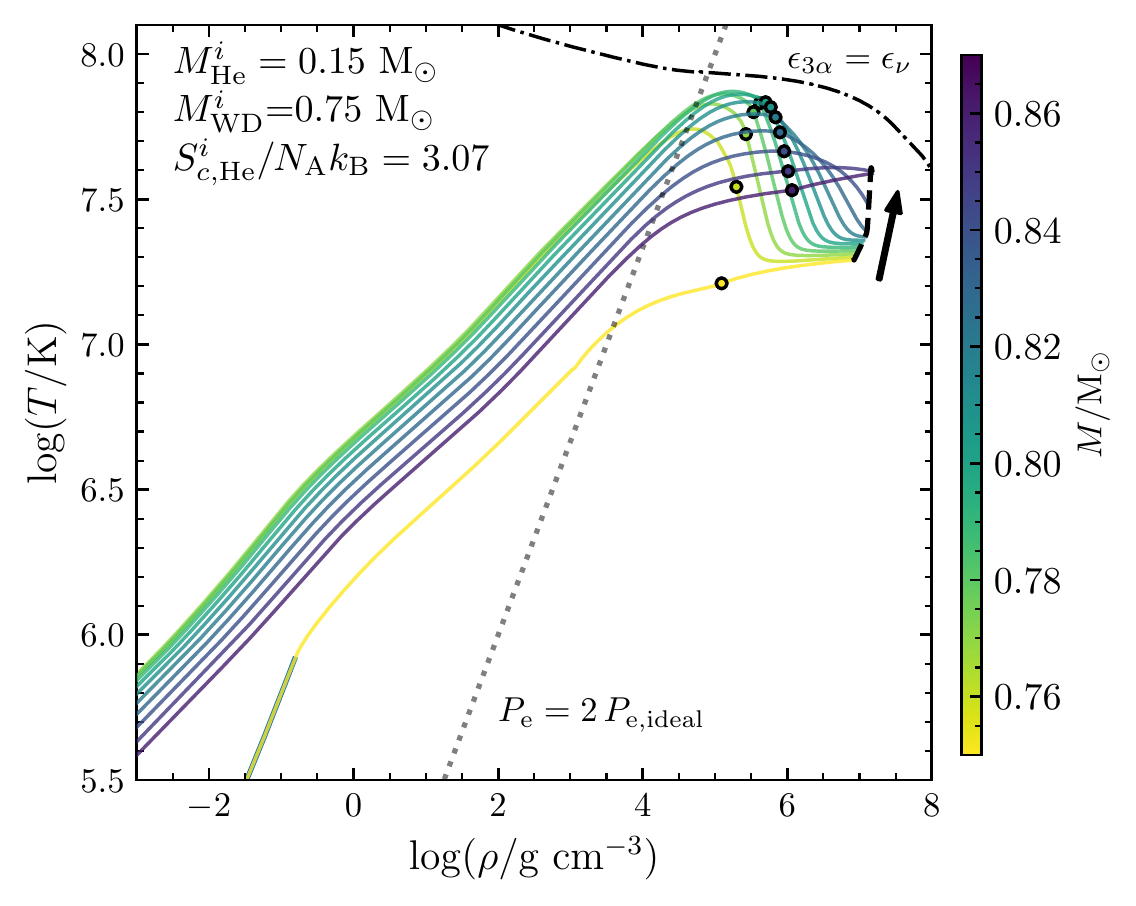}{0.5\textwidth}{(a)}
\fig{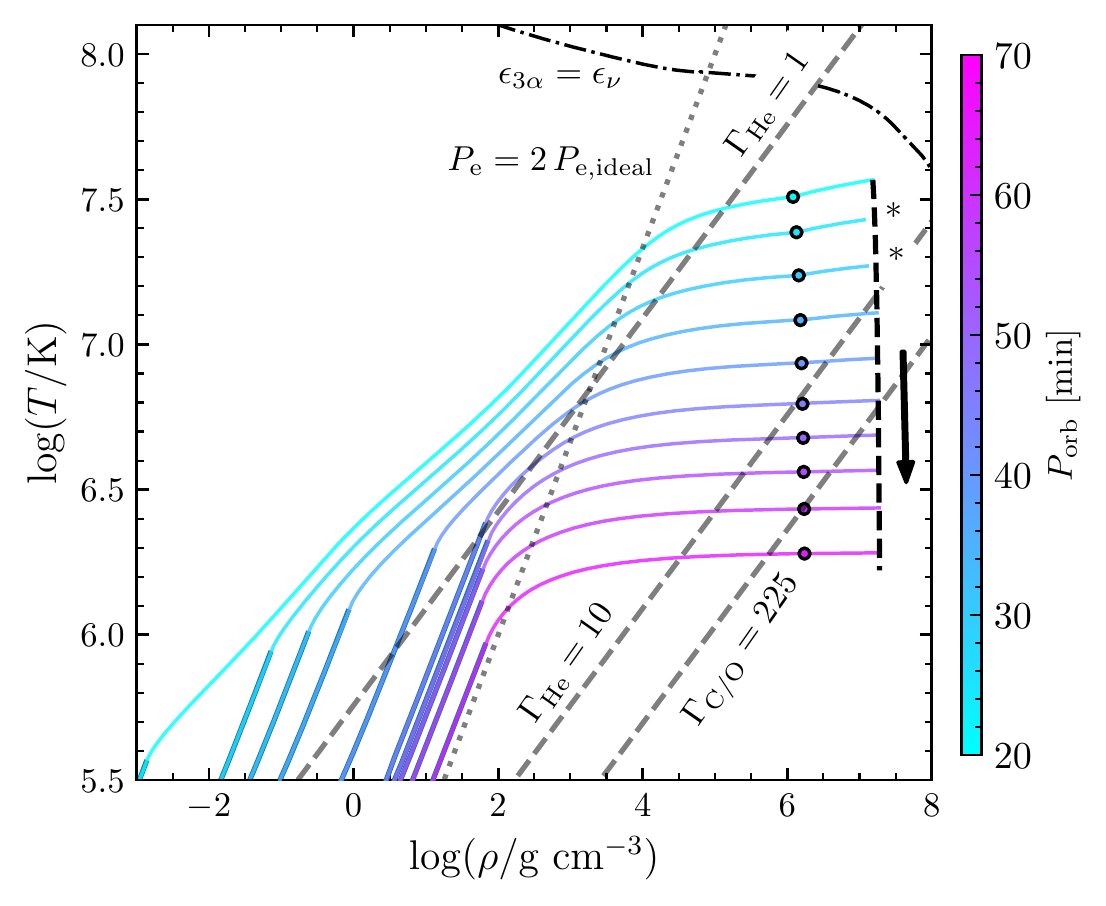}{0.49\textwidth}{(b)}
          }
\caption{ Density-temperature profiles of the WD accretor during core-heating (left panel, color-coded by $\mwd$) and core-cooling (right panel, color-coded by $\Porb$). The core temperature reaches a maximum at $\Porb \approx 20$ min and $\mwd\approx0.87 \, \msun$, after which the WD cools and crystallizes (at a Coulomb coupling parameter $\Gamma_{\mathrm{C/O}}=225$ for the core composition of the $\mesa$ model, as shown by the grey dashed line in the right panel). For each profile, we indicate the transition from the He envelope to the CO core by an open circle, and convection zone by overplotting a thick line. Core evolution is shown by the black dashed line and black arrow. We indicate the transition to degeneracy by where the electron pressure is twice the ideal gas electron pressure (dotted grey line), the start of helium burning by where the helium burning rate is balanced by thermal neutrino losses (dot-dashed black line), and two lines of constant Coulomb coupling parameter $\Gamma_{\mathrm{He}}$ for pure He (dashed grey lines). The two profiles with the labels * have $\Teff$ falling within the DBV instability strip \citep[$\num{22400} \leqslant \Teff/\mathrm{K} \leqslant \num{32000}$;][]{2019A&ARv..27....7C}. 
\label{fig:TRho_Evolution}
}
\end{figure*}

We now focus on the effects of mass transfer on the accretor properties, particularly $\Twd$. Given the small scatter in period-age relation over a range of $\iniThe$ and $\iniMwd$, we fix $\iniMhe = 0.15 \, \msun$, $\iniShe/\NA \kB = 3.07$, and study $\iniMwd = 0.75$ and $0.85 \, \msun$. 

Once settled onto the more massive CO WD, the accreted He releases heat while it is compressed deep in the envelope. At the initially high $\mdot \sim 10^{-8} - 10^{-7} \, \msunyr$, the rate of compression is higher than that of heat transport, resulting in a local temperature increase \citep{1982ApJ...253..798N,2004ApJ...600..390T}. This ``compressional heating'' creates a temperature inversion in the envelope, clearly seen in Panel (a) of Figure \ref{fig:TRho_Evolution}, as the sequence of temperature-density profiles of the accretor at various times (color-coded by accretor mass). Heat transport from the temperature peak occurs both towards the surface and into the colder core. The latter can be seen as a thermal front propagating to the center in Figure \ref{fig:TRho_Evolution}, which reaches the center in a few $10^{7}$ years after the start of mass transfer. Before then, the core evolves adiabatically. The $(1 \mathrm{-} 3) \times 10^{7}$ years for the core to become heated is consistent with the heat conduction time, $\tcond$, as derived in \citealt{2004ApJ...600..390T} (their equation A2). All models presented in this section avoided the occurrence of a helium flash, as the temperature of the helium layer never reached a value adequate to trigger a thermonuclear instability (see Figure \ref{fig:TRho_Evolution}). 
 
 When the age since mass transfer is comparable to $\tcond \approx (1 \mathrm{-} 3) \times 10^{7} $ yr, $\Twd$ reaches a maximum and the WD cools. 
 By then, $\Porb \gtrsim 20$ minutes and the donor has reached $0.03 \, \msun$. The mass accretion rate, $\mdot$, has dropped from peak by 1-2 orders-of-magnitude (see Figure \ref{fig:donor_time_comparison}), diminishing the importance of compressional heating and leading to a transition from core heating to core cooling \citep{B06}. Panel (a) of Figure \ref{fig:TRho_Evolution} shows that near peak $\Twd$, the temperature inversion gradually disappears as the core is heated due to the inward heat flow. We show the accretor profiles in the subsequent cooling phase in Panel (b) of Figure \ref{fig:TRho_Evolution}, color-coded by orbital period. As the WD further cools, the surface convection zone deepens and eventually crystallization starts when $\Gamma_{c} = 225$, at $\Porb \approx 50 $ min. 
 
\begin{figure}
\fig{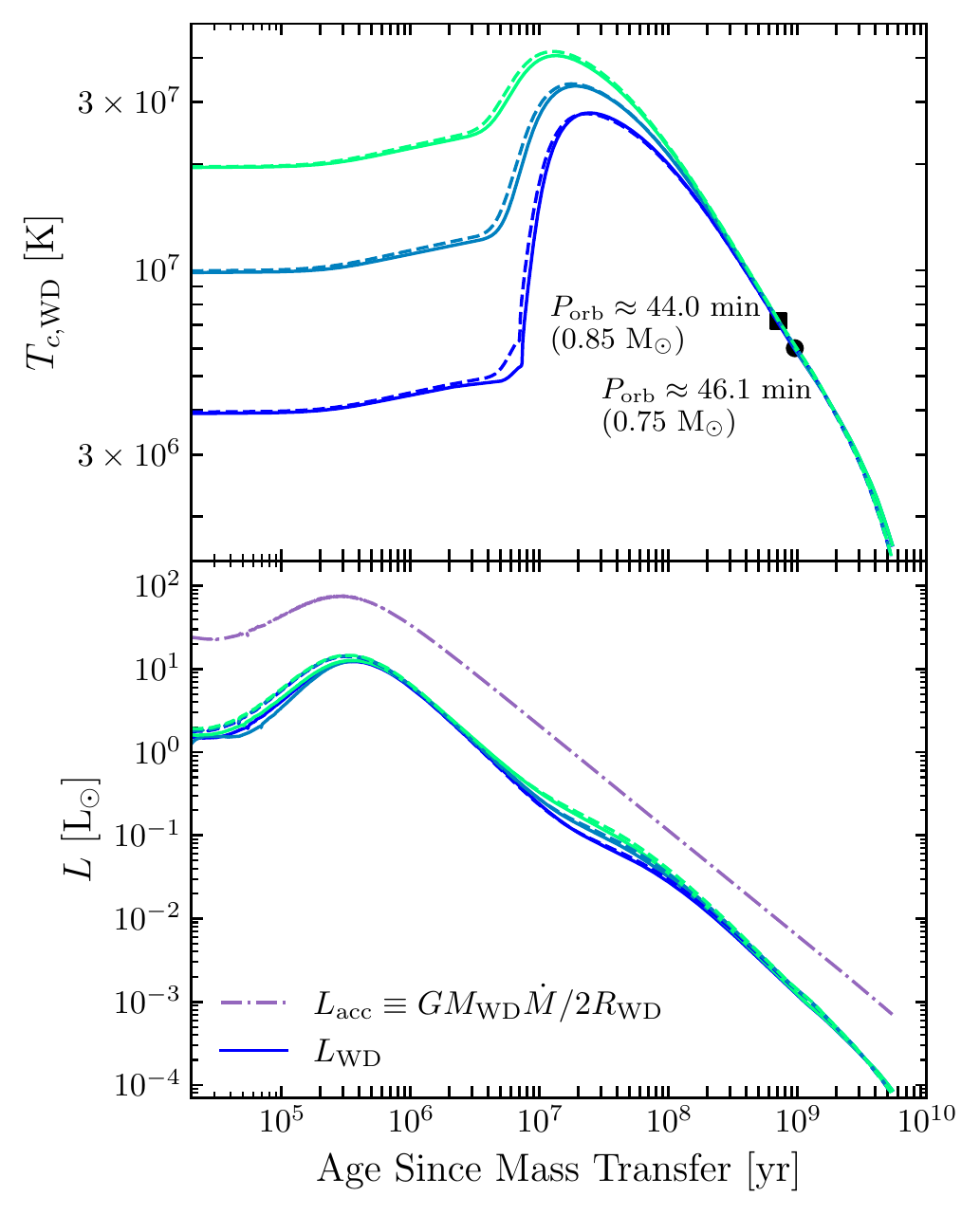}{ 0.49 \textwidth}{}
\caption{ Time evolution of the accretor core temperature, $\Twd$ (top panel), and accretor luminosity (bottom panel), since the start of mass transfer. The solid (dashed) lines show models with an initial accretor mass $\iniMwd=0.75 \, (0.85)\,\msun$, each set with a range of initial core temperatures for the accretor, $\iniTwd$ ($5 \times 10^{6}$ for deep-blue to $2 \times 10^{7}$ K for light green). We show where the models start to crystallize ($\Gamma_{c}=225$) by a circle (square) symbol for the $\iniMwd=0.75 \, (0.85) \, \msun$ models. We also show the accretion luminosity, $\Lacc \equiv G \mwd \mdot / 2 \Rwd$ (dot-dashed line). 
\label{fig:accretor_time_comparison}
}
\end{figure}

 The time evolution of $\Twd$ and $\Lwd$ is shown in Figure \ref{fig:accretor_time_comparison}. Consistent with Figure \ref{fig:TRho_Evolution}, the core first evolves adiabatically, then becomes significantly heated once the conduction front reaches the center at $10^{7} \, \mathrm{yrs}$, and cools thereafter. An initially cooler WD is heated to a lower peak $\Twd$ and hence shows a lower $\Lwd$ around peak $\Twd$. We note that, if a steadily accreting disk is present, its bolometric accretion luminosity $\Lacc \equiv G \mwd \mdot / 2 \Rwd$ would always dominate $\Lwd$. We discuss the challenges of differentiating between luminosity from accretion and from the cooling WD in the Appendix. 

Recent calculations by \cite{2020ApJ...899...46B} show a lower conductive opacity in WD envelopes, than the \cite{2007ApJ...661.1094C} conductive opacities which are used in typical WD cooling models \citep[e.g.,][]{2020ApJ...901...93B}. We incorporated this correction in all AM CVn models presented in this work. Compared to a model without this correction, the reheated accretor reaches a slightly lower peak central temperature, due to faster transport of heat out of the envelope. In the cooling phase, the accretor has a higher luminosity at the same central temperature (by $\approx 30 \%$) and cools faster. Equivalently, in well agreement with \cite{2020ApJ...899...46B}, this leads to a lower luminosity at a fixed cooling age. However, the difference in inferred cooling ages is much less than 1 Gyr, for a $0.9 \, \msun$ DB WD with $\Teff \approx 10^{4} \, \mathrm{K}$ (their Fig. 4), so its effect on our calculations is small.

\begin{figure}
\fig{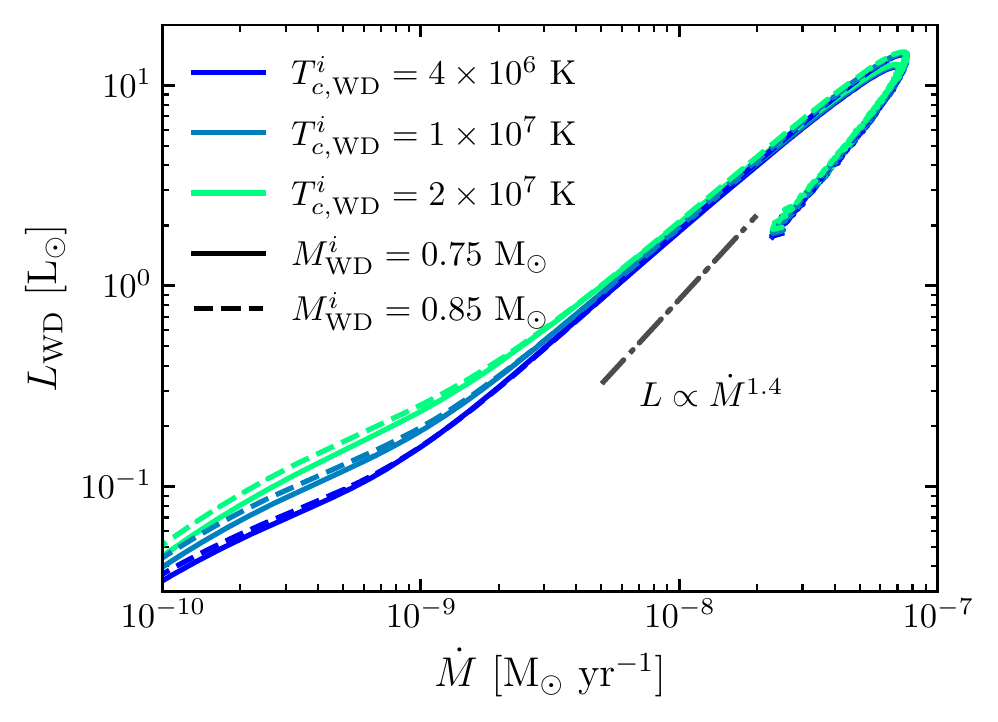}{ 0.49 \textwidth}{}
\caption{ Luminosity of the accretor, $\Lwd$, as a function of the mass transfer rate, $\mdot$, during the rapid accretion phase for the same set of models as in Figure \ref{fig:accretor_time_comparison}. We show $\Lwd \propto \mdot^{1.4}$ (dot-dashed grey line) as expected from \cite{B06} for comparison.
\label{fig:MdotL_comparison}}
\end{figure}

We compare our results with \cite{B06}, who modeled the thermal state of the accretor using quasistatic envelope methods. For early times, before the core gets heated, they found that the WD luminosity varies only with $\mdot$ and $\mwd$ as $\Lwd \propto \mdot^{1.4} \mwd^{-0.3}$ for $\mdot > 10^{-9} \msunyr$. We compare this expression with our $\mesa$ results in Figure \ref{fig:MdotL_comparison}. For $\mdot > 3 \times 10^{-9} \, \msunyr$, the $\mesa$ models show no dependence on $\iniTwd$ as expected \citep{B06}. Compressional heating dominates $\Lwd$ at early times. The power-law dependence of $\Lwd$ on $\mdot$ is slightly shallower than predicted. For $\mdot \lesssim 3 \times 10^{-9 } \, \msunyr$, the WD cooling luminosity starts to dominate, and an initially cooler WD is heated to a smaller peak $\Twd$ (see Figure \ref{fig:accretor_time_comparison}). Thus between $\mdot = 10^{-10} \, \msunyr$ and $3 \times 10^{-9} \, \msunyr$, a dependence of $\Lwd$ on $\iniTwd$ is seen, with a lower $\iniTwd$ giving a lower $\Lwd$ at fixed $\mdot$. 
 
\begin{figure}
\fig{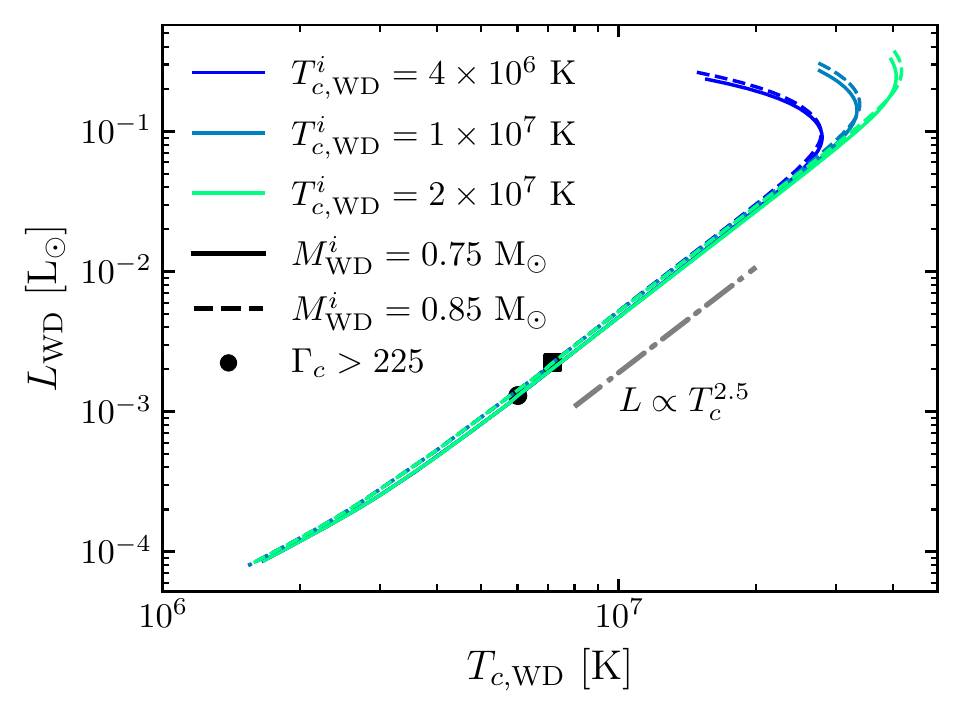}{ 0.49 \textwidth}{}
\caption{ Luminosity of the accretor, $\Lwd$, as a function of the core temperature of the accretor, $\Twd$, during the core-cooling phase. The models are the same as in Figure \ref{fig:accretor_time_comparison}. At late times, $\Lwd \propto \Tc^{2.5}$ as expected for cooling WDs, and is independent of $\iniTwd$ \citep{B06}. 
\label{fig:TcL_comparison}}
\end{figure}

\cite{B06} also predicted that at late times ($\Porb \gtrsim 30 \, \mathrm{min}$), the luminosity of the accretor should follow that of a cooling WD, $\Lwd \approx \lsun (\Tc/10^{8} \, \mathrm{K})^{2.5} (\mwd/0.6 \, \msun)$. Figure \ref{fig:TcL_comparison} shows excellent agreement between our models and the predicted relation. The $\Lwd$ tracks for the $\iniMwd=0.75 \, \msun$ models are roughly 0.89 times those for the $\iniMwd = 0.85 \, \msun$ models, consistent with the linear $\mwd$ dependence of $\Lwd$ and reflects the ratio $0.9/1.0 = 0.9$ in the final WD mass ($\mwd$ is nearly constant beyond $\Porb \gtrsim 20 \, \mathrm{min}$, see bottom panel of Fig \ref{fig:donor_time_comparison}). More importantly, at late times ($\Porb \gtrsim 30 \, \mathrm{min}$), the accretor evolves as a cooling WD independent of $\mdot$ and $\iniTwd$. Then, the binary age since mass transfer can be taken as a cooling age for the accretor, since the few $10^{7}$ yrs required to heat the WD is insignificant compared to the expected 4-5 Gyrs required for the binary to evolve to, say, $\Porb \approx 65 $ min (see Section \ref{sec:donor}). 

However, inferring the cooling ages of AM CVn accretors (see Section \ref{sec:comparison to observations}) is subject to the caveat that, compared to normal CO WDs, AM CVn accretors may have a significant fraction of their total mass in a thick He shell, if unstable He burning can be avoided (e.g., our $\iniMhe = 0.15 \, \msun $, $\iniMwd = 0.75 \, \msun $ model would have $q_{\mathrm{He}} = M_{\mathrm{He}}/\mwd = 0.15/0.9 \approx 0.17$ in the end). 
Since He has a different specific heat capacity than a C/O mixture, the cooling luminosity of an AM CVn accretor would be different than that of a CO WD of the same total mass, at a given cooling age. 

If we combine $ \Lwd = - \CV ( \diff \Tc/ \diff t )$, where $\CV$ is the total heat capacity, with $ \Lwd \propto \Tc^{5/2}$ \citep[e.g.,][]{2004ApJ...600..390T} and solve for the cooling time $\taucool$, then $\taucool \propto \CV \Lwd^{-3/5} $. 
As seen in Figure \ref{fig:TRho_Evolution}, at the ages $1 - 5 $ Gyrs that we are interested in, the C/O is near the crystallization limit and the He is approximately in the liquid state. Then we can approximate, for an AM CVn accretor that has a C/O core mass $M\CO$ and He shell mass $M\he$, $\CV \approx 3 ( \kB M\CO / \mu\CO m_{\mathrm{p}} ) + 2 ( \kB M\he / \mu\he m_{\mathrm{p}} )$, where $\mu\he = 4$, $\mu\CO \approx 14$, and the specific heat capacities for C/O and He are motivated by our $\mesa$ models \citep[see also,][]{2019MNRAS.490.5839B}. Similarly, for a CO WD of the same total mass $\Mtot = M\CO + M\he$, $\CV \approx 3 ( \kB \Mtot / \mu\CO m_{\mathrm{p}} )$. 
Hence, for a CO WD and an AM CVn accretor of the same total mass $\Mtot$ and cooling luminosity $\Lwd$, the ratio of their cooling ages is $ \taucoolAM / \taucoolCO \approx ( M\CO / \Mtot ) + (2/3)(\mu\CO/\mu\he)(M\he/\Mtot) $. For our fiducial case, $\iniMhe = 0.15 \, \msun$ and $\iniMwd = 0.75 \, \msun$, this ratio is about $1.22$. 

Therefore, when comparing the contours of constant age for CO WDs with thin He envelopes ($q_{\mathrm{He}} = 10^{-2}$) in Figure \ref{fig:SDSS_PS1_CMD} to observed data points, the AM CVn accretors may have older actual cooling ages than inferred, by tens of percent. Nonetheless, as we will show in Section \ref{sec:comparison to observations}, there exists a cooling age discrepancy of $\approx 3-4 $ Gyrs at $\Porb \approx 65 \, \mathrm{min}$ between theory and observations, so the corrections to the cooling age presented here are insignificant.

\begin{figure}
\fig{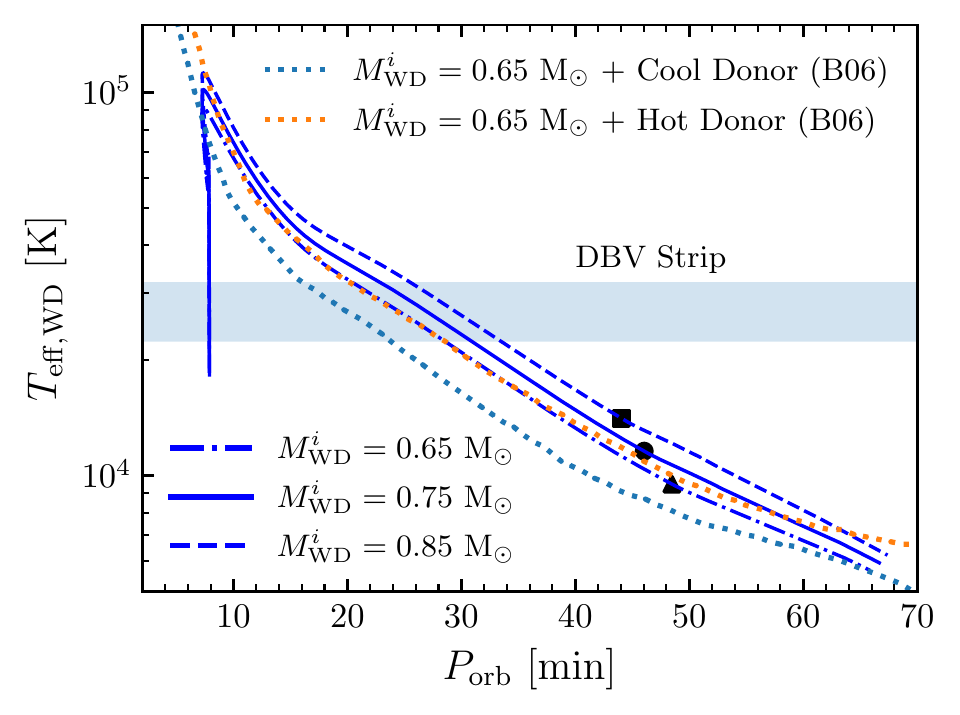}{ 0.49 \textwidth}{}
\caption{ Effective temperatures of the accretor, $\Teffwd$, as a function of orbital period, $\Porb$. Here we show our models with $\iniTwd=2 \times 10^{7}$ K and $\iniMwd = 0.65, 0.75 $ or $0.85 \, \msun$, the $\iniMwd = 0.65\, \msun$ with ``hot'' or ``cold'' donor models (dotted orange or light-blue lines) from \cite{B06}, and the DBV instability strip \citep[blue stripe; e.g.,][]{2019A&ARv..27....7C} for comparison. 
\label{fig:Porb_Teff}}
\end{figure}

Finally, the accretor effective temperature, $\Teffwd$, is shown as a function of $\Porb$ in Figure \ref{fig:Porb_Teff}. Models with $\iniMwd=0.65, 0.75 $ and $0.85 \, \msun$, all with $\iniTwd = 2 \times 10^{7}$ K and identical initial donor models, are shown. The WD starts cooling at $\approx 20 $ min, so subsequent evolution of $\Teffwd$ with $\Porb$ is set by the period-age relation of the binary. For comparison, we show the $\iniMwd=0.65$ models with a ``hot''/``cold'' donor from \cite{B06}. Our $\iniMwd=0.65$ accretor model initially tracks closely the ``hot'' donor model and later at long periods ($\Porb \gtrsim 60 \, \mathrm{min}$) the ``cold'' donor model. This is because our donor model starts with high-entropy and we allow for its subsequent cooling. The period-age relation then starts to deviate from the ``hot'' donor model which cools adiabatically, and to converge towards the ``cool'' donor model. We also show the DBV strip \citep[$\num{22400} \leqslant \Teff/\mathrm{K} \leqslant \num{32000}$; e.g.,][]{2019A&ARv..27....7C}, though no AM CVn accretors are presently known to be pulsating. Finding one would certainly enable a new probe of both the WD thermal state and thickness of the He layer.

\section{Comparison to Observations}
\label{sec:comparison to observations}

We showed that at $ \mdot \lesssim 3 \times 10^{-9} \, \msunyr $, the accretor luminosity is just that of a cooling WD. \textit{If} in addition we assume that, active accretion is not contaminating the data \citep{B06}, then combining a cooling WD model with the age-$\Porb$ relation yields a theoretical prediction for the luminosity of an AM CVn system at a given $\Porb$. We compare this prediction for the accretor to observed systems, and show that the observed systems appear $\approx 3 - 4 \, \mathrm{Gyrs}$ younger than predicted at $\Porb \approx 65 \, \mathrm{min}$.

To compare with observations, we compute absolute magnitudes from our $\mesa$ models. Given an effective temperature, $\Teff$, and surface gravity, $\log g$, from $\mesa$, we interpolate from the pure-helium model atmosphere table from the Montreal group \citep[][]{2011ApJ...737...28B,2018ApJ...863..184B,2020ApJ...901...93B}\footnote{\dataset[https://www.astro.umontreal.ca/$\sim$bergeron/CoolingModels/]{https://www.astro.umontreal.ca/~bergeron/CoolingModels/}} and obtain a bolometric correction for various photometric systems. 

Our V-band absolute magnitude, $\MV$, when plotted against $\Porb$, follows closely the ``cool'' donor track of \cite{B06}. However, we discovered that their $\MV$ is always lower by $\approx 0.7 \, \mathrm{mag}$ (i.e., brighter) than the $\MV$ we obtain using the Montreal bolometric corrections with the corresponding $\Teff$ and an assumed $\log g = 8.5$. For example, at $\Porb \approx 63 \, \mathrm{min}$, their ``cool'' donor model has $\Teffwd \approx 6000 \, \mathrm{K}$ which should give $\MV \approx 14.949$, but their $\MV$ reads $\approx 14.3$ instead. This explains why our $\MV-\Porb$ track agrees better with their ``cool'' donor track than with their ``hot'' donor track, despite better agreement in $\Teff-\Porb$ with the latter.

\begin{figure}
\fig{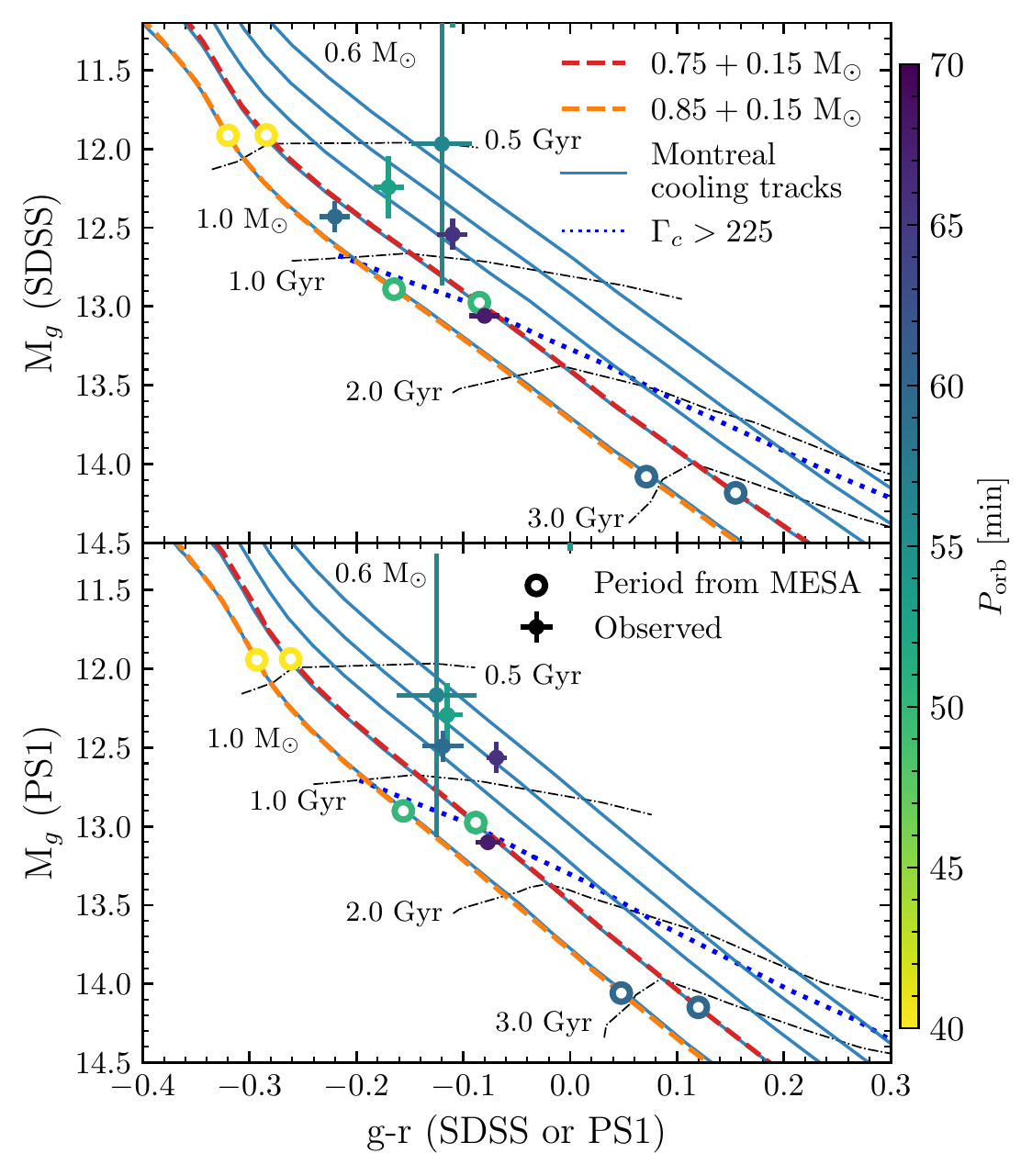}{ 0.49 \textwidth}{}
\caption{ Color-magnitude diagram (M$_{g}$ vs $g-r$) of the $\mesa$ models for SDSS (Pan-STARRS 1) bands in the top (bottom) panel. The red (orange) dashed line shows the $\mesa$ models with $\iniTwd=0.75 \, (0.85) \, \msun$, where we apply bolometric corrections from the DB atmosphere models from the Montreal group \citep{2011ApJ...737...28B,2018ApJ...863..184B,2020ApJ...901...93B}. We label where the $\mesa$ models reach a certain $\Porb$ with open circle symbols. For comparison we show their DB cooling models at $\mwd = 0.6 - 1.0 \, \msun$ (blue solid lines) and contours of constant cooling age (grey dot-dashed lines), and a contour of when $\mesa$ CO WD cooling models start crystallizing ($\Gamma_{c} > 225$; dotted blue line). The error bars show observed AM CVn systems \citep{Ramsay2018} with $\Porb > 50$ min, color-coded by $\Porb$. 
\label{fig:SDSS_PS1_CMD}}
\end{figure}

Regardless, as shown by \cite{Ramsay2018} (their Fig. 2), our $V$ or $g$-band absolute magnitudes, $\MV$ or $\Mg$, are lower than observed by 2 magnitudes at $\Porb \approx 60$ min. We show this discrepancy instead by comparing our $\mesa$ tracks on a color-magnitude diagram ($\Mg$ vs $g-r$ for SDSS and Pan-STARRS-1 filters; Fig \ref{fig:SDSS_PS1_CMD}), to observed AM CVn systems with $\Porb > 50 $ min, both color-coded by $\Porb$. By $\Porb \approx 20$ min, our $\iniMwd = 0.75 \, \msun$ (or $0.85 \, \msun$) accretor has accumulated most of the donor's mass, and starts to cool as a $\finalMwd = 0.9 \, \msun$ (or $1.0 \, \msun$) WD. This explains the good agreement of our $\mesa$ tracks even before $\Porb \approx 40$ min with the corresponding $0.9 \, \msun$ (or $1.0 \, \msun$) cooling tracks from \cite{2020ApJ...901...93B}. 

Nevertheless, even though the observed long-period systems lie close to the WD cooling tracks, they appear brighter and bluer than expected from our $\mesa$ tracks for the corresponding $\Porb$. \textsl{If} we interpret the observed source as dominated by the luminosity of the cooling WD, then their photometric cooling ages, as inferred from the CMD, are between 0.5 and 2 Gyr. In contrast, the $\mesa$ age-period relation (Sec \ref{sec:donor}) would give a cooling age of $4-5$ Gyr at $\Porb \approx 65$ min. 

\section{Conclusion}
\label{sec:conclusion}

In this work, we perform binary calculations with $\mesa$ for AM CVn binaries with a He WD donor. In Section \ref{sec:donor}, we evolved He WD donor models of various initial central specific entropies, $\iniShe$, and showed that the initial entropy and the subsequent thermal evolution of the donor dictates the mass transfer history and hence the age-period evolution of the binary. We find that at $ \Porb \approx 40 \, \mathrm{min} $, the donor starts to cool as its thermal timescale, $\tauth$, is comparable to or shorter than its mass transfer timescale, $\taum$, although this is subject to uncertainties in the surface opacity as described in Section \ref{sec:Donor Thermal Evolution Uncertainties Under Mass Transfer}. We then evolve the accretor along with the donor and the orbit in Section \ref{sec:heating}. At the initially high $\mdot$ near period minimum, the accretor is heated due to compressional heating. Later as $\mdot$ drops while the orbit widens, the accretor behaves simply as a cooling WD (at $\Porb \gtrsim 30 \, \mathrm{min}$). Our calculations show well agreement with the semi-analytic predictions of \cite{B06}, but are more accurate since we self-consistently consistent both binary components. 

Given that we theoretically expect that accretor is simply a cooling WD at $\Porb \gtrsim 30 \, \mathrm{min}$, we compute synthetic color-magnitude diagrams for our accretor models and compare with observations \citep[][]{Ramsay2018} in Section \ref{sec:comparison to observations}. We show that the observed systems, \textsl{if} interpreted as the cooling WD, appear much younger (in terms of cooling age) than expected at the corresponding $\Porb$ (by $\approx 3 - 4$ Gyrs at $\Porb \approx 65$ min). 
This may be attributed in part to uncertainties in the initial entropy distribution and subsequent cooling of the donor, which affects the age-period relation of our models. Better understanding of the donor cooling can be obtained by incorporating opacities of warm dense helium in $\mesa$. Another possibility is that the observed luminosities are contaminated by the accretion disk or boundary layer (discussed in the Appendix), although this appears unlikely in the case of Gaia14aae as revealed by eclipse modeling \citep[][]{2018MNRAS.476.1663G}. Finally, isolated WDs show evidence of a cooling delay \citep[e.g.,][]{Cheng2019,Kilic2020}, and we speculate that a similar phenomenon in the accretor may well explain the observed discrepancy. The recent discovery of 5 new eclipsing AM CVn systems with $\Porb = 35 - 62 \, \mathrm{min}$ by \cite{vanRoestel2021}, which constrains the luminosity contributions from different components and the degree of cooling of the donor, will help disentangle these various scenarios. 


\begin{acknowledgements}

We thank the referee for their constructive suggestions that have greatly improved our manuscript. 
We thank Evan Bauer and Josiah Schwab for helpful conversations about WD cooling and guidance on using $\mesa$, and the latter in addition for implementing the \cite{2020ApJ...899...46B} electron conductive opacity corrections in $\mesa$. We are grateful to Matthew Green, Thomas Kupfer, Tom Marsh, and Jan van Roestel, for valuable discussions about observations of AM CVn systems. We thank Sihao Cheng for making the python package $\texttt{WD\_models}$ publicly available.
This research was supported in part by the National Science Foundation through grant PHY-1748958, and  the Gordon and Betty Moore Foundation through grant GBMF5076. Use was made of computational facilities purchased with funds from the National Science Foundation (CNS-1725797) and administered by the Center for Scientific Computing (CSC). The CSC is supported by the California NanoSystems Institute and the Materials Research Science and Engineering Center (MRSEC; NSF DMR 1720256) at UC Santa Barbara.

\end{acknowledgements}

\software{%
\texttt{MESA} \citep[v12778;][]{2011ApJS..192....3P, 2013ApJS..208....4P, 2015ApJS..220...15P, 2018ApJS..234...34P, 2019ApJS..243...10P}, 
\texttt{ipython/jupyter} \citep{perez_2007_aa,kluyver_2016_aa},
\texttt{matplotlib} \citep{hunter_2007_aa},
\texttt{NumPy} \citep{numpy2020}, 
\texttt{SciPy} \citep{scipy2020}, 
\texttt{Astropy} \citep{astropy:2013,astropy:2018},
\texttt{Python} from \href{https://www.python.org}{python.org}
and 
\texttt{WebPlotDigitizer} \citep{Rohatgi2020}
.}

\clearpage

\appendix

\twocolumngrid

\section{Accretion Luminosity}
\label{sec:accretion luminosity}

We now address the possibility that observed AM CVn systems have contributions from the accretion disk or boundary layer. As the bottom panel of Figure \ref{fig:accretor_time_comparison} shows, if accretion is steady, the accretion luminosity, $\Lacc \equiv G \mwd \mdot / 2 \Rwd $, would be more than 5 times greater than the WD luminosity, $\Lwd$, at all times. 

\textsl{If} we assume steady-state accretion, and model the disk as a multicolor blackbody, we would expect the disk to dominate over the cooling WD in the V-band. We note that \cite{B06} proposed the opposite, because they assumed a uniform, averaged disk temperature defined by $\Lacc = S \sigma T^{4}$, where $S$ is the disk surface area. Their approach underestimated the temperature in the inner disk and hence the disk contribution to the $V$-band (again, assuming steady state). This implies we cannot expect \textsl{a priori} that the cooling WD dominates over the disk contribution in the V-band. 

However, the disk may not be in a steady state between 20 and 60 minutes \citep{Ramsay2018,2020ApJ...900L..37R,2021MNRAS.505..215R}, as it is expected to be thermally and viscously unstable if the mass transfer rate, $\mdot$, is between the limits $\mdotcrp$ and $\mdotcrm $ \citep[][their eqn. A2 for $Z=0.02$]{Kotko2012}, shown in Figure \ref{fig:pm_mdot_Porb}. 
The upper limit, $\mdotcrp$, is evaluated at the tidal radius $\Rtid = 0.6 a/(1+q)$, where $a$ is the binary separation and $q = \mhe/\mwd$ is the mass ratio. 
The lower limit, $\mdotcrm $, is evaluated at $\Rwd$, the accretor radius at the end of the simulation (essentially the radius of a $0.75 + 0.15 = 0.9 \, \msun$ WD). 

As noted by \cite{2015ApJ...803...19C}, \cite{Kotko2012} did not account for the zero-torque inner boundary condition in converting from $\Teff(R)$ to $\mdot(R)$ locally in the disk. This affects the lower stability condition $\mdotcrm$, since the global condition for the disk to be cold and stable is $\Teff(R) < \Teffcrm(R)$ for every radius $R$ within the disc. Because the disk effective temperature peaks at $(49/36) \Rwd$, with peak value given by $0.488 \Tstar$ where $\Tstar = 3 G \mwd \mdot / (8 \pi \Rwd^{3} \sigmasb )$, the lower stability condition should actually be $0.488 \Tstar < \Teffcrm(R=(49/36) \Rwd)$\footnote{The \citealt{Kotko2012} approach essentially takes $\Tstar < \Teffcrm(R=\Rwd)$ and solves for $\mdot$.}. Assuming $\Teffcrm \propto R^{\beta}$, where $\beta \approx -0.09$, and solving for $\mdot$, the actual $\mdotcrm$ should be a factor $(49/36)^{4 \beta} / 0.488^{4} \approx 15.8$ times $\mdotcrm(R=\Rwd)$. Thus, the green dashed line in the top panel of Figure \ref{fig:pm_mdot_Porb} should be 15.8 times higher if we account for the zero-torque inner boundary condition, and we would expect no disk outbursts for $\Porb \gtrsim 50$ min. However, this is to be taken with a grain of salt since AM CVn disk outbursts are yet to be fully understood \citep[e.g.,][]{Kotko2012,2020ApJ...900L..37R}.

Moreover, observations suggest an insignificant disk contribution for long-period AM CVn systems. Eclipse modeling of Gaia14aae suggests that the WD contribution is $\approx 80 \%$ \citep[][though a He WD donor is unlikely for this system]{2018MNRAS.476.1663G}. In addition, optical spectra and photometry of long-period systems can be fitted with a single blackbody of temperature $\Teff \approx 10^{4} \, \mathrm{K}$ \citep[e.g., SDSS J1137+4054, $\Porb=59.6$ min, and SDSS J1505+0659, $\Porb = 67.8$ min;][]{2014MNRAS.439.2848C}, whereas an accretion disk spectrum is expected to be flatter \citep[e.g.,][]{2009A&A...499..773N}. 

These observational constraints can be met by an optically thick boundary layer. A boundary layer can be consistent with eclipse modeling by virtue of its compact emitting area. Coincidentally, the ``accretion temperature'' $\Tacc$, defined by $\Lacc = 4 \pi \Rwd^{2} \Tacc^{4}$, gives the inferred blackbody temperature $10^{4} \, \mathrm{K}$. Note that, however, this is difficult from a theoretical point-of-view, because the boundary layer may be optically thin for $\mdot \lesssim 10^{-10} \, \msunyr$, and it may not spread over the entire WD surface which implies an effective temperature much higher than $\Tacc$ \citep[e.g., ][]{2004ApJ...610..977P}. 


\bibliography{main}{}
\bibliographystyle{aasjournal}



\end{document}